\documentclass[english,a4paper,pre,twocolumn,amsmath,amssymb,longbibliography]{revtex4-1}
\usepackage{graphicx,color,soul}

\newcommand{\dd}{\mathrm{d}}

\newcommand{\pd}[2]{\frac{\partial #1}{\partial #2}}

\newcommand{\IInt}[3]{\int_{#2}^{#3}\dd #1\;}

\renewcommand{\vec}[1]{\mathbf #1}

\newcommand{\al}{\alpha}

\newcommand{\lam}{\lambda}

\newcommand{\om}{\omega}

\newcommand{\x}{\vec r}

\newcommand{\kT}{k_\text{B}T}

\begin{document}

\title{Thermodynamic Approach to the Self-Diffusiophoresis of Colloidal Janus Particles}

\author{Thomas Speck}
\affiliation{Institut f\"ur Physik, Johannes Gutenberg-Universit\"at Mainz, Staudingerweg 7-9, 55128 Mainz, Germany}

\begin{abstract}
  Most available theoretical predictions for the self-diffusiophoretic motion of colloidal particles are based on the hydrodynamic thin boundary layer approximation in combination with a solvent body force due to a self-generated local solute gradient. This gradient is enforced through specifying boundary conditions, typically without accounting for the thermodynamic cost to maintain the gradient. Here we present an alternative thermodynamic approach that exploits a direct link between dynamics and entropy production: the local detailed balance condition. We study two cases: First, we revisit self-propulsion in a demixing binary solvent. At variance with a slip velocity, we find that propulsion is due to forces at the poles that are perpendicular to the particle surface. Second, for catalytic swimmers driven through liberating chemical free energy we recover previous expressions. In both cases we argue that propulsion is due to asymmetric dissipation and not simply due to an asymmetric concentration of molecular solutes.
\end{abstract}

\maketitle

%% ---- introduction ----

Colloidal particles that undergo not only thermal diffusion but directed motion have sparked substantial interest~\cite{kapral13,bech16,moran17,pour18}, both as synthetic microswimmers that allow to explore the mechanisms behind the collective dynamics of bacteria and cells~\cite{baue18}, and for their potential applications as microengines~\cite{sanchez14}. The prevalent experimental strategy is self-phoretic mechanisms that create a local gradient, \emph{e.g.}, of a molecular solute (diffusio or osmiophoresis)~\cite{hows07,butt12}, temperature (thermophoresis)~\cite{jian10,breg14}, or charged solutes (electrophoresis). But also sound waves~\cite{wang12}, electric fields~\cite{bric13,yan16}, and actuated body oscillations~\cite{dreyfus05} can be exploited to achieve locomotion.

Directed motion requires the dissipation of available thermal or chemical free energy and thus is a non-equilibrium phenomenon. Still, the prevalent theories for phoretic motion largely neglect this aspect and derive expressions for the speed from approximate solutions of the Stokes equation in the presence of a body force due to gradients~\cite{gole05,gole07,saba12,khair13,michelin14,wurg15,graaf15,popescu16,ibrahim17,lisicki18}. One can then determine the work necessary to power the Janus particle~\cite{saba12}, but it would be desirable to have expressions for the speed that make the dependence on the driving forces transparent. To this end, here we propose an alternative approach that makes use of the \emph{local detailed balance} condition~\cite{seif12}, which provides a direct link between dynamics and entropy production.

\begin{figure}[b!]
  \centering
  \includegraphics{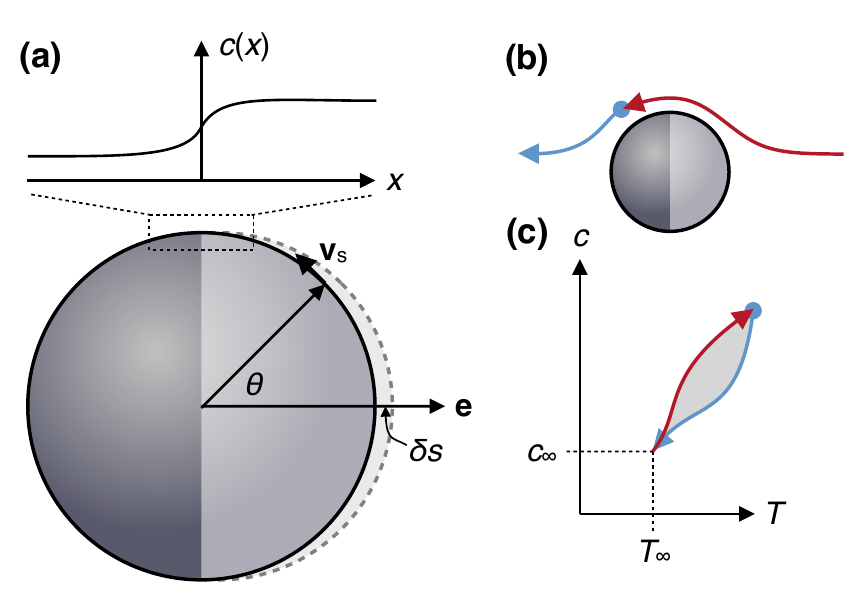}
  \caption{(a)~Axisymmetric spherical Janus particle with orientation $\vec e$. Indicated is a putative solvent slip velocity $\vec v_\text{s}(\theta)$ at polar angle $\theta$. The zoom sketches the surface concentration $c(x)$ of the molecular solute across the equator. (b)~Sketch of a solvent streamline close to the moving Janus particle in the comoving frame. (c)~Corresponding path of a solvent parcel in the temperature--concentration plane (with bulk values $T_\infty$, $c_\infty$) tracing a cycle. The gray area indicates the contribution to the work (excess dissipation) available to propel the Janus particle.}
  \label{fig:sketch}
\end{figure}

In the following, we focus on the self-diffusiophoresis of a single spherical particle caused by the gradient of a neutral molecular solute, cf. Fig.~\ref{fig:sketch}(a). The widely accepted picture is that interactions of the solutes with the particle surface lead to a pressure difference in the solvent, which drives a tangential flow characterized by the phoretic slip velocity $\vec v_\text{s}(\theta)$~\cite{ande89}. We restrict ourselves to axisymmetric particles with symmetry axis $\vec e$ and polar angle $\theta$ ($\theta=0$ corresponds to the front pole). In Stokes flow, force balance implies that the particle then is moving in the opposite direction with propulsion speed
\begin{equation}
  \vec v_\text{p} = -\frac{1}{2}\IInt{\theta}{0}{\pi} \sin\theta\vec v_\text{s}(\theta) = v_\text{p}\vec e.
\end{equation}
Since the solute gradient is due to spatially varying surface properties, it moves with the particle.

% ---- binary solvent ----

It is instructive to recapitulate the derivation of the slip velocity within the thin boundary layer approximation. Locally, the surface is assumed to be flat with the $x$ axis oriented along the flow and the $y$ axis corresponding to the surface normal. The Stokes equation for the solvent velocity field $\vec v(\x)$ reads
\begin{equation}
  0 = \eta\pd{^2v_x}{y^2} - \pd{p}{x} + f_x, \qquad
  0 = -\pd{p}{y} + f_y,
\end{equation}
with solvent viscosity $\eta$ and pressure $p$. In this boundary layer approximation, the normal velocity $v_y$ and its derivatives are assumed to be negligible, while the tangential velocity $v_x$ is assumed to vary rapidly along $y$ with $\partial_yv_x\gg\partial_x v_x$. The force density transmitted to the solvent is taken to be
\begin{equation}
  \label{eq:f}
  \vec f = -c\nabla\Phi,
\end{equation}
where $\Phi(\x)$ is the potential energy and $c(\x)$ the local concentration (or composition) of molecular solutes. Solving for the pressure and integrating twice yields the slip velocity $v_\text{s}(x)=v_x(x,y\to\infty)$. The flow field away from the particle is then solved using $v_\text{s}$ as boundary condition.

In Ref.~\citenum{wurg15}, this framework has been applied to a spherical colloidal Janus particle suspended in a solvent at temperature $T$. The solvent is a binary mixture, typically water and lutidine (the molecular solute), which has a lower critical temperature $T_\text{c}$ below which the mixture is homogeneous. The two hemispheres of the particle have different surface properties, geometrically defining two poles and an equator. The particle is illuminated, heating one hemisphere above $T_\text{c}$ so that the solvent demixes, leading to a concentration gradient of lutidine across the equator [Fig.~\ref{fig:sketch}(a)]. In experiments one indeed observes directed motion~\cite{butt12,schm18}. No directed motion is observed below $T_\text{c}$ (or in the absence of lutidine), which indicates that thermophoresis is not the dominant mechanism. Following Ref.~\citenum{wurg15}, the solvent slip velocity is predicted to become
\begin{equation}
  \label{eq:vs}
  v_\text{s}(x) \propto -\frac{k_\text{B}T}{\eta}\pd{c}{x},
\end{equation}
which only involves the derivative of the solute concentration $c(R,x)$ close to the surface. Here, $k_\text{B}$ is Boltzmann's constant and $R$ is the radius of the Janus particle.

However, a gradient of $c$ alone is not sufficient to drive solvent flow. For a counterexample, let us assume that one hemisphere is hydrophilic and the other hemisphere is hydrophobic, and that the conditions are such that the hydrophobic hemisphere is wetted by lutidine. Hence, even for $T<T_\text{c}$, the lutidine concentration $c(x)$ varies when crossing the equator. Clearly, the system is in thermal equilibrium, which implies that the solvent is at rest, at variance with Eq.~\eqref{eq:vs}. The reason is that the interfacial tension balances forces due to the gradient, and thus $\vec f=0$ everywhere demonstrating that Eq.~\eqref{eq:f} is insufficient.

In fact, a phoretic slip velocity is not necessary to predict directed motion. To derive the phoretic speed $v_\text{p}$, let us first assume that the colloidal particle moves with a constant speed $\vec u=u\vec e$ along its orientation. We assume that within the comoving frame of reference, the composition and temperature profiles are stationary. The heat dissipated in a small volume element due to the solvent flow equals the work,
\begin{equation}
  \delta\dot q(\x) = (\nabla\psi)\cdot\vec v,
\end{equation}
where $\psi(c,T)$ is the free energy density depending on the local concentration and temperature. The solvent speed can be written as $\vec v(\x)=u\vec n(\x)$ since in the Stokes regime it is proportional to $u$. The excess entropy production rate caused by moving the colloidal particle is
\begin{equation}
  \label{eq:ep}
  \dot S = \IInt{^3\x}{\mathcal V}{} \frac{\delta\dot q(\x)}{T(\x)}.
\end{equation}
The integral is over the space $\mathcal V$ occupied by the solvent excluding the particle. This entropy is produced in addition to the dissipation due to the heat flow driven by the temperature gradient.

Due to noise, the particle is not moving at a fixed speed. On the time scale $\delta t$, we assume that the particle undergoes discrete jumps with forward rate $k_+$ and backward rate $k_-$. Appealing to a time scale separation between slow particle motion and the fast response of the solvent, we assume a piece-wise ``steady state'' with speed $u$ during each jump. We now employ a result from stochastic thermodynamics, local detailed balance, that relates both rates with the entropy production
\begin{equation}
  \frac{k_+}{k_-} = e^{\dot S\delta t/k_\text{B}}
\end{equation}
due to moving the particle a distance $\delta s=u\delta t$. We parametrize rates as $k_+=k_0e^{\al\dot S\delta t/k_\text{B}}$ and $k_-=k_0e^{(\al-1)\dot S\delta t/k_\text{B}}$ with some $\al$. The probability $P(s,t)$ for the particle to have traveled a length $s$ (the arc length of the trajectory for which the tangent is proportional to the particle orientation) up to time $t$ obeys the master equation
\begin{equation}
  \pd{P}{t} = k_+P(s-\delta s,t) + k_-P(s+\delta s,t) - (k_++k_-)P(s,t).
\end{equation}
Expanding the probability
\begin{equation}
  P(s\pm\delta s,t) = P(s,t) \pm \pd{P}{s}\delta s + \frac{1}{2}\pd{^2P}{s^2}(\delta s)^2 + \cdots
\end{equation}
and rates
\begin{gather}
  k_+ = k_0[1+\al\dot S\delta t/k_\text{B}+\cdots], \\
  k_- = k_0[1+(\al-1)\dot S\delta t/k_\text{B}+\cdots],
\end{gather}
to order $(\delta s)^2$ we find the drift-diffusion equation
\begin{equation}
  \label{eq:diff}
  \pd{P}{t} = -v_\text{p}\pd{P}{s} + D\pd{^2P}{s^2}
\end{equation}
with (constant) diffusion coefficient $D=k_0(\delta s)^2$ and propulsion speed
\begin{equation}
  \label{eq:vp:bin}
  v_\text{p} = D\IInt{^3\x}{\mathcal V}{} \frac{(\nabla\psi)\cdot\vec n}{k_\text{B}T(\x)}.
\end{equation}
Both expressions are independent of $\al$. Moreover, to leading order the diffusion coefficient $D$ is independent of the entropy production and thus describes thermal diffusion, albeit in the presence of spatially inhomogeneous solvent temperature and viscosity~\cite{ring10}.

To make quantitative predictions one would of course need to know (in the comoving frame) the temperature profile $T(\x)$, the concentration profile $c(\x)$, and the free energy density $\psi(c,T)$. But even without explicit expressions one can draw some general insights from Eq.~\eqref{eq:vp:bin} into what drives the propulsion. To this end, let us consider a fluid parcel of the solvent moving along a streamline (with tangent proportional to $\vec n$) close to the particle. Along this streamline, both the composition $c$ and the temperature $T$ change as sketched in Fig.~\ref{fig:sketch}(b). Approaching the particle, the temperature rises from its bulk value $T_\infty$ and the parcel is heated, extracting energy from the temperature gradient. At the same time, the concentration of the solute increases. At some point this is reversed and the parcel is cooled until it reaches bulk values again. For non-vanishing net dissipation, the two paths in the $T$-$c$ plane need to be different [cf. Fig.~\ref{fig:sketch}(b)], otherwise the cooling parcel would dump the same free energy it had taken up before. A \emph{moving} solvent parcel thus acts as an engine, transferring \emph{excess} heat from hot to cold through changing the concentration (akin to changing the volume in a conventional engine). This excess heat is spent to propel the particle according to Eq.~\eqref{eq:vp:bin}.

Both the temperature and the concentration profile decay with the distance to the particle. This implies that $\nabla\psi\sim\vec e_r$ predominately points along the radial direction. Hence, the largest contribution to Eq.~\eqref{eq:vp:bin} actually comes from the poles with forces perpendicular to the surface (but $\pd{c}{x}\sim0$) and not the equator (where $\pd{c}{x}$ is largest but $\vec n$ and $\nabla\psi$ are perpendicular). Indeed, in agreement with our results, a recent numerical study~\cite{sami15} found evidence that self-propelled motion in a binary solvent is not to be attributed to a slip velocity parallel to the surface. They show that no significant slip builds up and that the flow pattern remains similar to that of Stokes flow past a viscous droplet.

For symmetric flow $n_r(-\theta)=n_r(\theta)$ at least two variables are necessary (here the temperature and concentration) to obtain net dissipation and thus propulsion. Thermophoresis~\cite{jian10,breg14} occurs in a self-generated temperature gradient alone. In this case one needs to break also the symmetry of the flow in order to obtain propulsion.

% ---- chemical reaction ----

Self-propulsion of Janus colloids can also be achieved through harnessing spatially non-uniform chemical reactions, in particular through covering one hemisphere with a catalyst while the other hemisphere is chemically inert~\cite{hows07,ebbens12}. Most theoretical approaches aim to determine the ensuing non-isotropic concentration profiles of reactants and products (specified by the boundary conditions) and again relate those to a slip velocity on the particle's surface~\cite{gole05,gole07,graaf15,popescu16,ibrahim17} (or an effective osmotic force~\cite{cord08}). Aspects like the relevance of advection~\cite{khair13} and the inherent stochasticity of the chemical reactions~\cite{gasp17,gaspard18} have also been discussed.

For a single reactant molecule on the particle surface, the chemical conversion to product occurs with rates $\om_+$ (forward) and $\om_-$ (backward) obeying the detailed balance condition
\begin{equation}
  \frac{\om_+}{\om_-} = e^{\Delta\mu/\kT}
\end{equation}
with $\Delta\mu$ the difference of chemical potential between product and reactant driving the reaction. The dissipation on a small surface patch with area $\delta A$ is $\delta\dot q_\text{tot}=\dot N\Delta\mu$ with flux
\begin{equation}
  \dot N(\theta) = (\om_+-\om_-)K(\theta)c(R,\theta)\ell\delta A.
\end{equation}
Here, $c(R,\theta)$ is the concentration of reactant molecules on the particle surface and $\ell$ is the thickness of the thin surface layer within which molecules undergo the reaction (not to be confused with the hydrodynamic boundary layer). Finally,
\begin{equation}
  K(\theta) = \begin{cases}
    1, & 0 \leqslant \theta < \pi/2, \\
    0, & \pi/2 \leqslant \theta < \pi,
  \end{cases}
\end{equation}
is the coverage function describing the catalytic patch.

The heat rate $\delta\dot q_\text{tot}$ corresponds to the total dissipation. Similar to the heat flow needed to maintain the thermal gradient, this is not what determines the directed motion. As before, we need to identify the \emph{excess} dissipation due to moving the particle with speed $u$. To this end, we require the current of reactants, which reads
\begin{equation}
  \vec j = -D_0\nabla c + u\vec n c
\end{equation}
with $\nabla\cdot\vec j=0$ in bulk and boundary condition $j_r(R,\theta)=-\dot N/\delta A$. Here, $D_0$ is the diffusion coefficient of the reactant molecules and we employ the radial solvent flow $un_r(R+\ell,\theta)$ at the edge of the surface layer injecting (or removing) reactant molecules. Expanding $c=c_0+c_\text{ex}$, where $c_0$ is the solution for $u=0$, we neglect terms $u c_\text{ex}$ and $D_0\partial_rc_\text{ex}$ to find the excess flux $\dot N_\text{ex}\approx-u n_rc_0K(\theta)\delta A$. The excess heat thus reads $\delta\dot q=\dot N_\text{ex}\Delta\mu$. Following the same steps as above, we obtain the same drift-diffusion equation~\eqref{eq:diff} but now with propulsion speed
\begin{equation}
  \label{eq:vp:chem}
  v_\text{p} = -D\frac{\Delta\mu}{\kT}\IInt{^2\x}{\delta\mathcal V}{} n_r(\theta)c_0(R,\theta)K(\theta)
\end{equation}
integrating over the particle surface.

Excess reactants are convected by the solvent flow. We assume that this flow is still given by the Stokes flow around a no-slip sphere,
\begin{equation}
  n_r \approx \cos\theta\left[1-\frac{3R}{2(R+\ell)}+\frac{R^3}{2(R+\ell)^3}\right] \approx \frac{3\ell^2}{2R^2}\cos\theta,
\end{equation}
where we expand to lowest order of $\ell/R\ll 1$. Inserting this result into Eq.~\eqref{eq:vp:chem} and using the Stokes-Einstein relation $D=\kT/(6\pi\eta R)$ we obtain
\begin{equation}
  v_\text{p} = -\frac{\kT}{\eta R}\lam^2 \IInt{\theta}{0}{\pi}\sin\theta\cos\theta c_0(R,\theta)K(\theta)
\end{equation}
with length
\begin{equation}
  \lam^2 = \frac{1}{2}\ell^2\frac{\Delta\mu}{\kT}.
\end{equation}
We thus recover the same form for $v_\text{p}$ as boundary layer approaches~\cite{gole05,ebbens12}. At variance with the established explanation, however, we find that directed motion is not due to the concentration gradient \emph{per se} but due to an asymmetric dissipation on the hemispheres.

Directed motion requires $\Delta\mu\neq 0$ and an anisotropic $c(R,\theta)$. The latter condition is not necessarily fulfilled for uniform spherical particles with $K=1$, which, although in non-equilibrium, might not undergo directed motion. This has been demonstrated for uniform ion exchange particles, the interactions of which can be captured by an effective conservative potential~\cite{niu17}. Interestingly, even uniform particles might cause a spontaneous symmetry breaking of $c$ provided the timescales of solvent and concentration relaxation do not separate anymore~\cite{michelin13}.

%% ---- conclusion ----

To conclude, we have presented a thermodynamic approach linking the stochastic dynamics of phoretic Janus particles with excess entropy production through the local detailed balance condition. Here, we have studied neutral solute molecules, with a non-uniform concentration either due to reversible demixing or a surface reaction. In both cases we predict directed motion without resorting to a hydrodynamic slip velocity. Heating of one hemisphere and a difference of chemical potential, respectively, generate a non-equilibrium (approximately steady) state. For directed motion to occur, fluctuations of the particle position need to cause excess dissipation, for which the steady driving is a prerequisite. Catalytic swimmers based on redox reactions also involve electric currents that influence the propulsion~\cite{brow14,brown17}, and it would be interesting to extend our approach to this situation. Finally, we hope that the insights into a single particle presented here will help to understand the collective behavior of interacting chemically powered microswimmers~\cite{zott14,lieb15,yan16a,huang16}.

%% ---- bibliography ----

\end{document}